\documentstyle[preprint,aps,graphics]{revtex}
\tightenlines
\title {Coherent processing of a light pulse stored in a medium of four-level atoms}
\author{A. Raczy\'nski$^{1}$
\footnote{email: raczyn@phys.uni.torun.pl}, J. Zaremba$^{1}$ and
S. Zieli\'nska-Kaniasty$^{2}$}
\address{$^{1}$Instytut Fizyki, Uniwersytet Miko\l aja Kopernika,
       ul.Grudzi\c{a}dzka 5,
       87-100 Toru\'n, Poland,\\$^{2}$ Instytut Matematyki i Fizyki,
       Akademia Techniczo-Rolnicza, Al. Prof. S. Kaliskiego 7, 85-796
       Bydgoszcz, Poland.}
\begin{document}

\maketitle

\begin{abstract}
It is demonstrated that the properties of light stored in a
four-level atomic system can be modified by an additional control
interaction present during the storage stage. By choosing the
pulse area of this interaction one can in particular continuously
switch between two channels into which light is released.
\\
 PACS numbers: 42.50.Gy, 03.67.-a
\end{abstract}
\newpage
In recent years a number of nonlinear optical phenomena in weak
fields have been intensively investigated. They are connected with
a coherent excitation of an atomic medium, the optical properties
of which can then be drastically modified. An important example is
the electromagnetically induced transparency \cite{w1}, possibly
additionally controlled in time, which is manifested as light
slow-down or even its storage and a controlled release
\cite{w2,w3,w4,w5,w6,w7}. The simplest realization of those
effects occurs in atomic systems with three active states in a
$\Lambda$ configuration. Extending such a configuration by adding
a coherently coupled fourth level opens new possibilities of an
external control of such processes \cite{w8,w9,w10,w11} . In our
previous paper \cite{w12} we have shown that in a double $\Lambda$
system it is possible to change the light frequency of the stored
light or even to release two pulses of different frequencies by
applying two control fields, properly chosen and delayed in time.

Light storing in the form of atomic coherences joins the
advantages of the efficiency of light as an information carrier
and of an atomic medium as an information store. Thus it might be
a question of a practical importance how to modify in a controlled
way the properties of the released light by processing the atomic
medium during the storage stage. In the present work we
investigate two possibilities of controlling the released pulse or
pulses by modifying the atomic coherence due to the stopped light.
In the case (a) the lower, initially empty state of a typical
$\Lambda$ system is additionally coupled to a fourth state by
another laser. In the case (b) the initially occupied state of the
$\Lambda$ system is coupled by some kind of an effective
interaction (e.g., magnetic or two-photon electric coupling) with
another state of the same parity. We show that the Rabi
oscillations due to the new interactions modify in a coherent way
the properties of the released light.

We consider a quasi one-dimensional medium of four-level atoms
with three lower metastable states $|b>$ and $|c>$ and $|d>$ and
an upper state $|a>$ (Fig.\ref{fig1}). The position of an atom is
described by the variable $z$, which is considered continuous. The
states $b,a$  and $c$ constitute a typical $\Lambda$ system with
the weak signal field 1 and a strong control field 2. In the case
(a) the state $c$ is additionally coupled with another state $d$
by a laser field 4. In the case (b) the state $a$ is coupled with
a fourth state $d$ by a weak signal field 3 while the states $b$
and $d$ are coupled by some effective coupling 4. The interaction
Hamiltonian in the case (a) is $V=-\hat{d}\sum_{j=1,2,4}
\epsilon_{j}\cos\phi_{j}$, with $\phi_{j}=\omega_{j} t-k_{j} z$,
$\epsilon_{j}=\epsilon_{j}(z,t)$ being slowly varying envelopes
and $\hat{d}$ - the dipole moment operator; we have assumed that
all the fields have the same linear polarization. In the case (b)
the Hamiltonian reads $V=-\hat{d}\sum_{j=1,2,3}
\epsilon_{j}\cos\phi_{j}+ [(iU)|b><d|-(iU)|d><b|] \cos \phi_{4}$,
where for numerical reasons we have made a physically
insignificant assumption that the matrix element $iU$ of the
effective interaction 4 is imaginary. The matrix elements of the
dipole moment $d_{1}=(\hat{d})_{ab}$, $d_{2}=(\hat{d})_{ac}$,
$d_{3}=(\hat{d})_{ad}$, $d_{4}=(\hat{d})_{cd}$ are taken real.
Resonant conditions concerning all the couplings are assumed.

The evolution equation $i\hbar \dot{\rho}=[H,\rho]$ for the
density matrix $\rho=\rho(z,t)$ for an atom at position $z$, after
making the rotating-wave approximation, transforming-off the
rapidly oscillating factors: $\rho_{ab}=\sigma_{ab}
\exp(-i\phi_{1})$, $\rho_{ac}=\sigma_{ac} \exp(-i\phi_{2})$,
$\rho_{bc}=\sigma_{bc} \exp[i(\phi_{1}-\phi_{2})]$,
$\rho_{db}=\sigma_{db} \exp(-i\phi_{4})$, $\rho_{dc}=\sigma_{dc}
\exp[i(\phi_{2}-\phi_{3})]$, $\rho_{ad}=\sigma_{ad}
\exp[i(\phi_{3})]$, $\rho_{ii}=\sigma_{ii}$, and after adding
relaxation terms describing the spontaneous emission within the
system, takes the form in the more complicated case (b)

\begin{eqnarray}
 i \dot{\sigma}_{aa}&=&
-\frac{1}{2\hbar}\epsilon_{1}d_{1}(\sigma_{ba}-\sigma_{ab})
+\frac{1}{2}\Omega_{2}(\sigma_{ca}-\sigma_{ac})
-\frac{1}{2\hbar}\epsilon_{3}d_{3}(\sigma_{da}-\sigma_{ad})
-i(\Gamma^{a}_{b}+\Gamma^{a}_{c}+\Gamma^{a}_{d})
\sigma_{aa},\nonumber\\
 i \dot{\sigma}_{bb}&=&
 -\frac{1}{2\hbar}\epsilon_{1}d_{1}(\sigma_{ab}-\sigma_{ba})
 +\frac{1}{2\hbar} iU (\sigma_{bd}+\sigma_{db})
+i \Gamma^{a}_{b} \sigma_{aa} ,\nonumber\\
 i \dot{\sigma}_{cc}&=&
\frac{1}{2}\Omega_{2}(\sigma_{ac}-\sigma_{ca})
 +i\Gamma^{a}_{c} \sigma_{aa},\nonumber\\
 i \dot{\sigma}_{dd}&=&
-\frac{1}{2\hbar}\epsilon_{3}d_{3}(\sigma_{ad}-\sigma_{da})
-\frac{1}{2\hbar} iU (\sigma_{bd}+\sigma_{db})+i \Gamma^{a}_{d}
\sigma_{aa},\nonumber\\
 i \dot{\sigma}_{ab}&=&-\frac{1}{2\hbar}\epsilon_{1} d_{1} (\sigma_{bb}-
 \sigma_{aa})+\frac{1}{2}\Omega_{2}\sigma_{cb}
 -\frac{1}{2\hbar}\epsilon_{3}d_{3}\sigma_{db}
+\frac{1}{2\hbar}iU\sigma_{ad}
 - \frac{i}{2}(\Gamma^{a}_{b}+\Gamma^{a}_{c}+\Gamma^{a}_{d})
 \sigma_{ab},\\
 i \dot{\sigma}_{ac}&=&-\frac{1}{2\hbar}\epsilon_{1} d_{1}
 \sigma_{bc}+
\frac{1}{2} \Omega_{2}(\sigma_{cc}-\sigma_{aa})
 - \frac{1}{2\hbar} \epsilon_{3}d_{3}\sigma_{dc}-
 \frac{i}{2}(\Gamma^{a}_{b}+\Gamma^{a}_{c}+\Gamma^{a}_{d})
 \sigma_{ac},\nonumber\\
 i  \dot{\sigma}_{ad}&=&-\frac{1}{2\hbar}\epsilon_{1} d_{1}
 \sigma_{bd}
+\frac{1}{2}\Omega_{2}\sigma_{cd}
 -\frac{1}{2\hbar} \epsilon_{3}d_{3}(\sigma_{dd}-\sigma_{aa})
-\frac{1}{2\hbar}iU\sigma_{ab}
  -\frac{i}{2}(\Gamma^{a}_{b}+\Gamma^{a}_{c}+\Gamma^{a}_{d})
 \sigma_{ad},\nonumber\\
 i \dot{\sigma}_{bc}&=&-\frac{1}{2\hbar}\epsilon_{1} d_{1}
 \sigma_{ac}
-\frac{1}{2} \Omega_{2}\sigma_{ba}
 +\frac{1}{2\hbar}iU\sigma_{dc},\nonumber\\
 i \dot{\sigma}_{bd}&=&
-\frac{1}{2\hbar}\epsilon_{1}d_{1}\sigma_{ad}
+\frac{1}{2\hbar}\epsilon_{3}d_{3}\sigma_{ba} + \frac{1}{2\hbar}
iU(\sigma_{dd}-\sigma_{bb}),\nonumber\\
 i \dot{\sigma}_{cd}&=&
\frac{1}{2} \Omega_{2}\sigma_{ad}
+\frac{1}{2\hbar}\epsilon_{3}d_{3}\sigma_{ca}- \frac{1}{2}
iU\sigma_{cb},\nonumber
\end{eqnarray}
where $\Gamma^{a}_{b}$ is the decay rate of the state $|a>$ to
$|b>$, etc., and $\Omega_{2}=-\epsilon_{2}d_{2}/\hbar$ is the Rabi
frequency corresponding to the driving field 2.

The corresponding equations for the case (a) are obtained from the
above set (1) by interchanging the indices $b\leftrightarrow c$,
$1\leftrightarrow 2$, by setting $\epsilon_{3}=0$, $d_{3}=0$,
$\Gamma^{a}_{d}=0$, by replacing the effective coupling $iU$ by
$=-\epsilon_{4} d_{4}$ (when multiplied by $\sigma_{dc}$,
$\sigma_{da}$, $\sigma_{ac}$, $\sigma_{db}$, $\sigma_{bc}$,
$\sigma_{cc}$ and $\sigma_{dd}$) and by changing the sign of
$\omega_{4}$.

The propagation equations for the signal field 1 (in the case (a))
and for both signal fields (1) and (3) (in the case (b)), in the
slowly varying envelope approximation and in the conditions of the
resonance read
\begin{eqnarray}
\frac{\partial \epsilon_{1}}{\partial z}+\frac{1}{c}
\frac{\partial \epsilon_{1}}{\partial t}=-i N
d_{1}\frac{\omega_{1}}{\epsilon_{0}c} \sigma_{ba},\nonumber\\
\frac{\partial \epsilon_{3}}{\partial z}+\frac{1}{c}
\frac{\partial \epsilon_{3}}{\partial t}=-i N
d_{3}\frac{\omega_{3}}{\epsilon_{0}c} \sigma_{da},
\end{eqnarray}
where $N$ is the atom density and $\epsilon_{0}$ is the vacuum
electric permittivity. Similarly as in earlier papers, we have
neglected propagation effects for the driving fields, i.e. we take
$\epsilon_{2,4}=\epsilon_{2,4}(t)$.

Eqs (1) and (2) have been solved numerically in the moving window
frame of reference: $t'=t-z/c,z'=z$. Switching the driving field 2
on and/or off was modeled by a hyperbolic tangent, while the
additional pulse 4 was taken rectangular. The initial probe pulse
was taken as the sine square shape
\\$\epsilon_{1}(0,t)=\epsilon_{10}
\sin^{2}[\pi(t-\tau_{1})/(\tau_{2}-\tau_{1})]\Theta(t-\tau_{1})
\Theta(\tau_{2}-t)$,
while the initial condition for the atomic part was
$\sigma_{bb}(z,0)=1$, with other matrix elements equal to zero.

We have performed model computations for data being of realistic
orders of magnitude, however without making attempt to imitate any
real atom. The atomic energies were $E_{a}$= -0.10 a.u.,
$E_{b}=$-0.20 a.u. $E_{c}=$-0.18 a.u. with $E_{d}=$ -0.22 a.u (in
the case (a)) and $E_{d}=E_{b}+10^{-7}$ a.u. in the case (b) (the
latter value is of order of a magnetic energy splitting). The
relaxation rates for the spontaneous emission from the level
$E_{a}$ to $E_{b}$, $E_{c}$ and in the case (b) also to $E_{d}$
were taken equal to $2.4\times 10^{-9}$ a.u., from which the
dipole moments have been calculated. The dipole moment for the
electric transition $c\leftrightarrow d$ was taken $-2.74\times
10^{-1}$ a.u., which corresponded to a negligible width of the
level $E_{c}$. The length of the atomic sample was
$2.5\times10^{7}$ a.u. (1.3 mm) in the case (a) and $3\times
10^{7}$ a.u. (1.6 mm) in the case (b) and its density $3\times
10^{-13} $ a.u. $(2\times 10^{12}$ cm$^{-3}$). The initial signal
pulse length was $10^{11}$ a.u. (2.4 $\mu$s) and
$\epsilon_{10}=10^{-10}$ a.u. (which corresponded to the power
density of 3.5$\times10^{-4}$ Wcm$^{-2}$); the maximum value of
the amplitude of the control field 2 was $1.2 \times10^{-9}$ a.u.
(50 mWcm$^{-2}$). The values of the $b-d$ effective coupling $U$
in the case (b) were of order of $10^{-10}$ a.u. while in the case
(a) we took a coupling with $\epsilon_{4}=2\times 10^{-9}$ a.u.

In Fig.\ref{fig2} we show the released part of the pulse 1 as a
function of the local time $t'$ for different values of the area
of the pulse 4 (case(a)). The pulse can be lowered, completely
damped or its sign reversed depending on the final phase of the
Rabi oscillations between the levels c and d. Of course the final
results do not depend on particular time instants of switching the
interaction 4 on and off, provided that the pulse arrived after
the signal pulse 1 has been stored and before the release stage
has started. The presence of the Rabi oscillations becomes clearly
visible in the situation in which the control pulses 2 and 4
partially overlap. In this case, with the latter pulse being now
by an order of magnitude stronger than before, the restored pulse
is constructed of parts freed in those intervals of the Rabi
period in which the coherence $\sigma_{bc}$ differs significantly
from zero. The Rabi oscillations between the levels c and d are
thus imposed on the leaving signal pulse (see Fig.\ref{fig3}).

The Rabi oscillations due to the additional control field may be
used not only to destroy in a reversible way the atomic coherence
$\sigma_{ab}$ necessary to release the pulse 1 (case (a)) but also
to create a new coherence $\sigma_{ad}$ which can be converted
into a new pulse 3 (case (b)). In Fig.\ref{fig4} we show the
shapes of the two signal pulses 1 and 3 for different values of
the area of the control pulse 4, switched on and off in the
storage stage. If the area is a multiple of $\pi$ only the pulse 1
is released, with its sign being changed in the case of an odd
multiple. If the area is an odd multiple of $\frac{\pi}{2}$ only
the pulse 3 appears, alternatively with a changed sign. For pulse
areas being not a particular multiple of $\frac{\pi}{2}$ both
pulses 1 and 3 are released, their heights being under control.

As in the previous papers the problem can be analyzed in terms of
dark state polaritons. Such an analysis allows one to describe the
whole process of light storing in a single $\Lambda$ system in
terms of a shape preserving solution of the Maxwell-Bloch
equations (Eqs (1,2)), the components of which, i.e. the signal
field and the atomic coherence, adiabatically turn one into
another. In the case of a four-level system the evolution could
not in general be fully adiabatic, which means that bright-state
polaritons must appear at some stage of the process and are later
damped \cite{w12}. Thus the dark-state polaritons at the initial
and final stages are not identical.

The approach of Ref. \cite{w12} generalized in our case (b) leads
to the following results. One can attempt to solve Eqs (1) and (2)
perturbatively (as concerns signal fields), in an adiabatic and
relaxationless approximation. The stage of light stopping occurs
as in the case of a three-level system: the polariton solution
\begin{equation}
\Psi=\frac{\Omega_{2}\epsilon_{1}+\frac{2\omega_{1}N
d_{1}}{\epsilon_{0}}\sigma_{bc}}
{\sqrt{\Omega_{2}^{2}+\frac{2\omega_{1}Nd_{1}^{2}}
{\hbar\epsilon_{0}}}}[-\frac{d_{2}}{|d_{2}|}]
\end{equation}
describes an adiabatic conversion of the pulse 1 into the
coherence $\sigma_{bc}$ and the sign correction guarantees that
for large $|\Omega_{2}|$ we get $\psi=\epsilon_{1}$ (we assume
that $\epsilon_{2}>0$). After pulse stopping, say at time instant
$t_{1}$, the control pulse 4 is switched on and is present up to
the time instant $t_{2}$. As a consequence the density matrix
evolves and at $t=t_{2}$ we obtain in the case (b)
$\sigma_{bb}=\cos^{2}\theta$, $\sigma_{dd}=\sin^{2} \theta$,
$\sigma_{bd}=-\sin\theta \cos\theta$,
$\sigma_{bc}=\sigma_{bc}(t_{1}) \cos\theta$,
$\sigma_{dc}=-\sigma_{bc}(t_{1})\sin\theta$, where
$\theta=\frac{U(t_{2}-t_{1})}{2\hbar}$ is the pulse area.

At time instant $t_{3}$ ($t_{3}>t_{2}$) the control field 2 is
switched on in order to release the trapped pulse. In the assumed
approximations $\sigma_{bb}$, $\sigma_{dd}$ and $\sigma_{bd}$ do
not change any more. The pulses 1 and 3 satisfy the equations
\begin{equation}
(\frac{\partial}{\partial t}+c\frac{\partial}{\partial z})
\epsilon_{j}=\frac{1}{\Omega_{2}}\frac{\partial}{\partial t}
\frac{1}{\Omega_{2}} \sum_{k} M_{jk} \epsilon_{k},
\end{equation}
where $j,k=1,3$ and
$M_{11}=-\frac{2N\omega_{1}d_{1}^{2}}{\epsilon_{0}
\hbar}\cos^{2}\theta$,
$M_{13}=\frac{2N\omega_{1}d_{1}d_{3}}{\epsilon_{0}
\hbar}\sin\theta\cos\theta$,
$M_{31}=\frac{2N\omega_{3}d_{3}d_{1}}{\epsilon_{0}
\hbar}\sin\theta\cos\theta$,
$M_{33}=-\frac{2N\omega_{3}d_{3}^{2}}{\epsilon_{0}\hbar }\sin^{2}
\theta$.

Eqs (4) can be decoupled by a linear transformation. One of the
solutions can be shown, similarly as in previous papers, to be a
shape-preserving solution traveling with a time-dependent velocity
\begin{equation}
v(t)=c\frac{1}{1+\frac{2N(d_{1}^{2}\omega_{1}\cos^{2}\theta
+d_{3}^{2}\omega_{3}\sin^{2}\theta)}{\hbar\epsilon_{0}\Omega_{2}^{2}}}.
\end{equation}
(The other solution is zero due to the initial conditions at
$t=t_{2}$.) The polariton, being a combination of two fields and
two coherences, has the form
\begin{eqnarray}
 \Psi=\sqrt{d_{1}^{2}\omega_{1}\cos^{2}\theta+d_{3}^{2}\omega_{3}
 \sin^{2}\theta} \frac{\sqrt{\omega_{1}})}
 {\sqrt{1+\frac{2N}{\epsilon_{0}\hbar \Omega_{2}^{2}}
 (d_{1}^{2}\omega_{1}\cos^{2}\theta+d_{3}^{2}\omega_{3}\sin^{2}\theta)}}
\frac{d_{1}}{|d_{1}|} \nonumber\\
 \times
[\frac{d_{1}\epsilon_{1}\cos\theta-d_{3}\epsilon_{3}\sin\theta}
 {d_{1}^{2}\omega_{1}\cos^{2}\theta+d_{3}^{2}\omega_{3}\sin^{2}\theta}
 +\frac{2N}{\epsilon_{0}\Omega_{2}}(\sigma_{bc}\cos\theta
 -\sigma_{dc}\sin\theta)].
\end{eqnarray}
The solution (6) has been normalized so that it is equal to the
solution (3) at $t=t_{2}$, i.e.
$\Psi(t_{2})=\sqrt{\frac{2N\hbar\omega_{1}}{\epsilon_{0}}}
[\sigma_{bc}(t_{2})\cos\theta-\sigma_{dc}(t_{2})\sin\theta]$.
However, the final form of the polariton is, again for large
$\epsilon_{2}$, a combination of the signal fields with the
coefficients different from $\cos\theta$ and $\sin\theta$, except
in the case of $d_{1}=d_{3}$ and $\omega_{1}=\omega_{3}$. In
particular for $\theta=-\frac{\pi}{2}$ one finds that
$\Psi\rightarrow \sqrt{\frac{\omega_{1}}{\omega{3}}}
\epsilon_{3}\frac{d_{1}d_{3}}{|d_{1}d_{3}|}$ instead of
$\epsilon_{3}\frac{d_{1}d_{3}}{|d_{1}d_{3}|}$. As described in
detail in our previous paper \cite{w12}, this means that the
evolution cannot be fully adiabatic and bright state polaritons
(which are later damped) must be invoked. In the case (a) the
nonadiabatic element of the evolution is even more conspicuous:
only the part of the transformed coherence, namely that
proportional to $\cos\theta$ is active in the release stage and
turns adiabatically, being a shape-preserving solution, into
$\cos\theta \epsilon_{1}$. The other part of the excitation,
namely that proportional to $\sin\theta$, "survives" the release
stage inside the medium unless relaxations in the states $b$ and
$d$ are taken into account.

In summary, we have demonstrated that a modification of the atomic
coherence due to a stopped light pulse can be used as a new way of
changing the properties of the released light. One can in
particular release two pulses of different frequencies or
polarizations, with their envelopes being regulated in a
continuous way. This may serve as a kind of a switch which allows
one e.g., to continuously steer the information by sending it to
particular channels or to temporarily hide it.

 \acknowledgments{This work is a part of a program of the
National Laboratory of AMO Physics in Toru\'n, Poland}

\newpage
\noindent

\begin{figure}
\caption{Level and coupling schemes; the indices 1 and 3 refer to
signal fields and 2 and 4 - to control fields.} \label{fig1}
\end{figure}

\begin{figure}
\caption{The field amplitude of the released pulse 1 as a function
of the local time $t'$ for different pulse areas of the control
field 4 in the case (a): curve 1: 0, curve 2: $\frac{\pi}{6}$,
curve 3: $\frac{\pi}{4}$, curve 4: $\frac{\pi}{3}$, curve 5:
$\frac{\pi}{2}$, curve 6: $\frac{3\pi}{4}$, curve 7: $\pi$,}
\label{fig2}
\end{figure}

\begin{figure}
\caption{The field amplitude of the released pulse 1 as a function
of the local time $t'$ in the case (a) for overlapping pulses 2
and 4: curve 1: part of the pulse transmitted before light
storing, curve 1a: the released pulse in the absence of the
additional coupling field, curve 1b: the released pulse in the
presence of the additional coupling field, curve 2: the control
field 2, curve 4: the additional coupling field 4. The values of
the fields 2 and 4 have been reduced by the factors of 40 and 20,
respectively.  } \label{fig3}
\end{figure}

\begin{figure}
\caption{The amplitudes (in $10^{-11}$ a.u.) of the signal fields
1: full line, and 3: dashed line,  as functions of the local time
$t'$ (in $10^{11}$ a.u.) in the case (b) for different values of
the pulse area of the interaction 4.} \label{fig4}
\end{figure}

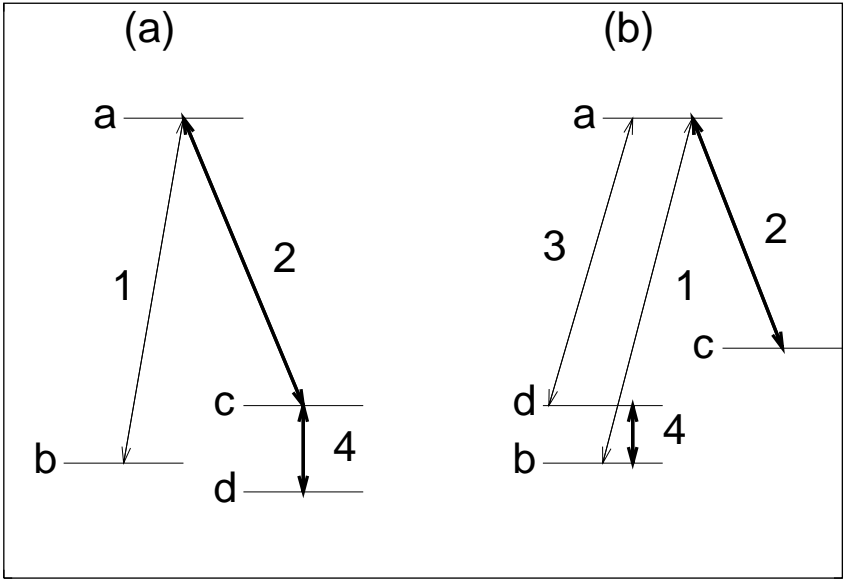
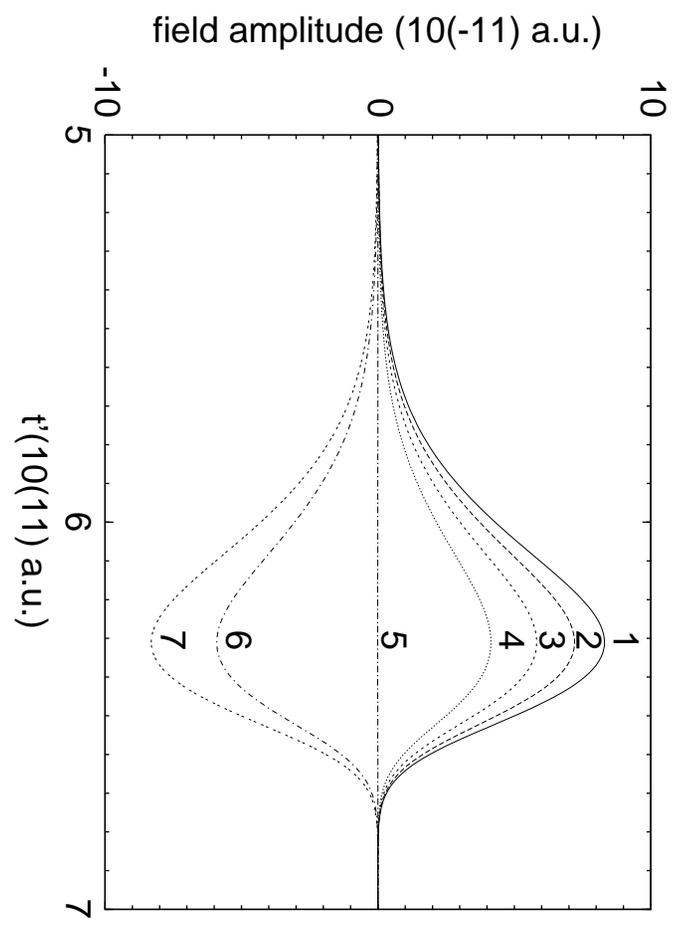
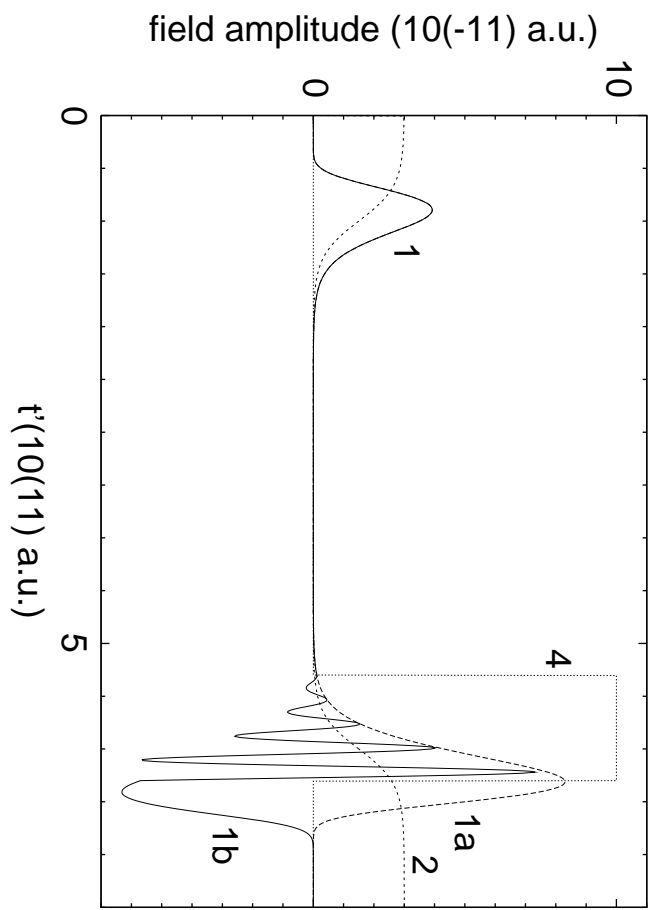
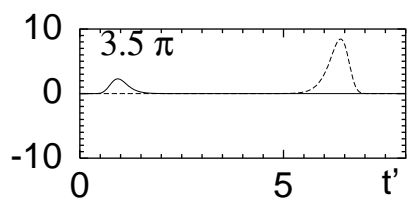
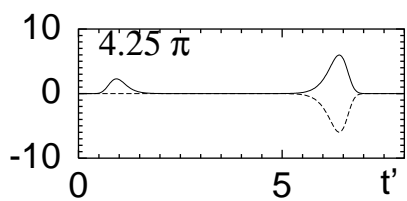
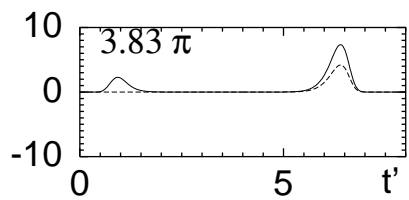
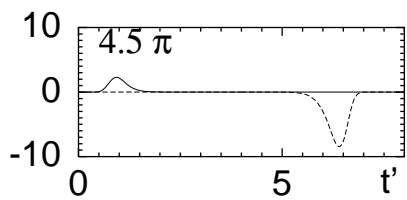
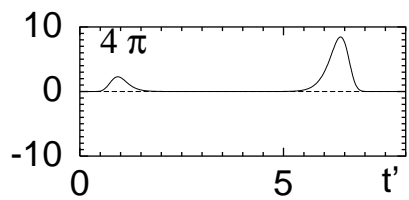
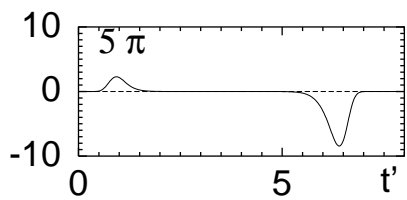
\end{document}